\documentclass[aps,prd,twocolumn,twoside,superscriptaddress,floatfix, nofootinbib]{revtex4-1}
\pdfoutput=1
\usepackage{graphicx}
\usepackage{color}
\usepackage{hyperref}
\usepackage{float}
\usepackage{aas_macros}
\usepackage{ifthen,amsmath,amssymb, bm}
\usepackage{multirow}
\usepackage{comment}
\usepackage[normalem]{ulem}

\graphicspath{{Figures/}}

\newcommand{\beq} {\begin{equation}}
\newcommand{\eeq} {\end{equation}}
\newcommand{\bal} {\begin{aligned}}
\newcommand{\eal} {\end{aligned}}

\newcommand{\barr}{\begin{eqnarray}}
\newcommand{\earr}{\end{eqnarray}}

\begin{document}

\title{Extragalactic CO emission lines in the CMB experiments: a forgotten signal and a foreground}

\author{Abhishek S. Maniyar}
\email{amaniyar@stanford.edu}
\affiliation{Kavli Institute for Particle Astrophysics and Cosmology, Stanford University, 452 Lomita Mall, Stanford, CA 94305, USA}
\affiliation{SLAC National Accelerator Laboratory, 2575 Sand Hill Road, Menlo Park, CA 94025, USA}
\affiliation{Center for Cosmology and Particle Physics, Department of Physics, New York University, New York, NY 10003, USA}

\author{Athanasia Gkogkou}
\affiliation{Aix Marseille Univ, CNRS, CNES, LAM, Marseille, France}

\author{William R. Coulton}
\affiliation{Center for Computational Astrophysics, Flatiron Institute, 162 5th Avenue, New York, NY 10010, USA}

\author{Zack Li 
}
\affiliation{Canadian Institute for Theoretical Astrophysics, University of Toronto, Toronto, ON, Canada M5S 3H8}

\author{Guilaine Lagache}
\affiliation{Aix Marseille Univ, CNRS, CNES, LAM, Marseille, France}

\author{Anthony R. Pullen}
\affiliation{Center for Cosmology and Particle Physics, Department of Physics, New York University, New York, NY 10003, USA}

\date{\today}

\begin{abstract}
High resolution cosmic microwave background (CMB) experiments have allowed us to precisely measure the CMB temperature power spectrum down to very small scales (multipole $\ell \sim 3000$). Such measurements at multiple frequencies enable separating the primary CMB anisotropies with other signals like CMB lensing, thermal and kinematic Sunyaev-Zel'dovich effects (tSZ and kSZ), and cosmic infrared background (CIB). In this paper, we explore another signal of interest at these frequencies that should be present in the CMB maps: extragalactic CO molecular rotational line emissions, which are the most widely used tracers of molecular gas in the line intensity mapping experiments. Using the SIDES simulations adopted for top hat bandpasses at 150 and 220 GHz, we show that the cross-correlation of the CIB with CO lines has a contribution similar to the CIB-tSZ correlation and the kSZ power, thereby contributing a non-negligible amount to the total power at these scales. This signal, therefore, may significantly impact the recently reported $\geq 3\sigma$ detection of the kSZ power spectrum from the South Pole Telescope (SPT) collaboration, as the contribution of the CO lines is not considered in such analyses. Our results also provide a new way of measuring the CO power spectrum in cross-correlation with the CIB. 
Finally, these results show that the CO emissions present in the CMB maps will have to be accounted for in all the CMB auto-power spectrum and cross-correlation studies involving a LSS tracer. 
\end{abstract}

\maketitle

\section{Introduction} \label{sec:intro}
Observations of the cosmic microwave background (CMB) continue to help us understand the properties of the Universe. A distinct advantage of CMB observations is that different CMB anisotropies contain information coming from different epochs of the Universe. For example, the primary CMB anisotropies which show up on large angular scales in the CMB power spectrum ($\ell < 2000$) precisely constrain properties of the Universe around redshift $z \sim 1100$ \cite{Planck_cosmo_2020}. On the other hand, secondary anisotropies like CMB lensing, thermal and kinematic Sunyaev-Zel'dovich effects (tSZ and kSZ) are tracers of the low redshift Universe. They show up on relatively smaller angular scales ($\ell > 2000$) and, as a result, are hard to measure and disentangle. 
Additional sources of emission, such as the cosmic infrared background (CIB), sometimes dominate over these anisotropies and make these measurements even more difficult. However, due to the wealth of crucial information they contain, precisely measuring these small-scale CMB anisotropies is a major goal of current and future CMB experiments. This requires careful modeling of all sources of anisotropies and emissions present at such scales.

Fortunately, the ever-increasing sensitivity and angular resolution of CMB experiments like Planck \cite{Plancktsz_2016}, ACT \cite{Dunkley_2013}, SPT  \cite{George_2015}, SO \cite{Ade_19}, CMB-S4 \cite{Abazajian_16} have brought us closer to this goal. Although foregrounds like the CIB contaminate CMB maps near the frequencies where the CMB dominates ($\sim 150$ GHz),
they have a distinct spectral shape with respect to the CMB. As a result, an effective way to constrain different components of the CMB map 
is to perform the measurement across different frequencies and utilize this distinct spectral dependence. These experiments have done exactly this 
by using their multi-frequency data to successfully separate different individual contributions to the CMB power spectrum at these small scales. In fact, for the first time, the latest results from the SPT experiment measured the evasive kSZ power spectrum at $\geq 3\sigma$ \cite{Reichardt_2021}.
Measuring the kSZ power spectrum is challenging due to its small amplitude and the presence of other sources like the tSZ, and CIB \cite{Dunkley_2013, George_2015, Reichardt_2021}. 

The kSZ signal contains crucial information about the reionization history of the Universe, which is otherwise extremely difficult to obtain. The SPT's achievement, therefore, has paved a way for future, more sensitive measurements of the kSZ power spectrum and, as a result, a better understanding of the epoch of reionization. In light of the upcoming more sensitive CMB data, it is imperative to carefully scrutinize the current analysis methods to perform a comprehensive study by considering all possible sources of power at these frequencies. \cite{Reichardt_2021} perform a thorough analysis of the small-scale SPT data at 95, 150, and 220\,GHz. Apart from the primary CMB, they consider the following templates and fit for them: tSZ, CIB, CIB-tSZ cross-correlation, radio-source emission, galactic dust, and kSZ. 

In this paper, we explore another signal of interest at these frequencies: bright CO molecular rotational line emissions. They have garnered interest especially in the field of line intensity mapping (LIM). LIM experiments aim to make a statistical observation of the aggregate emission from many unresolved galaxies rather than aiming to image individual galaxies directly (see \cite{Kovetz_2017, Bernal_2022} for a review). In LIM studies, CO lines are the most widely used tracers of molecular gas, which is a crucial interstellar medium component. These rotational emission lines are among the brightest in the galaxy spectra and occur at the ladder of frequencies $\nu_{J \rightarrow J-1} = J \times 115.27$ GHz for the $J \rightarrow J-1$ transitions. 
Different CO transitions from different redshifts will be observed in CMB bands ($\sim 150$ GHz). As first pointed out by \cite{Righi_2008}, the CO lines should be a source of foreground for the CMB observations. However, they also point out that the amplitude of the CO power spectrum goes down with the spectral resolution of the instrument i.e. the amplitude goes down with lower values of $\Delta \nu/\nu_{\rm obs}$, where $\Delta \nu$ is the bandwidth of the instrument at the observed frequency $\nu_{\rm obs}$. As a result, for a SPT-like instrument with $\Delta \nu \approx 50$\,GHz, 
the amplitude of different CO lines acting as a foreground should be small \cite{Righi_2008}. For example, according to the model used by \cite{Righi_2008} for different CO lines, with $\Delta \nu/\nu_{\rm obs} = 0.2$ at $70$\,GHz, all the CO lines have their respective power spectrum amplitudes $\mathcal{D}_\ell = \ell (\ell + 1)C_\ell/2\pi \:  < 0.1 \, \mu {\rm K}^2$ at $\ell = 3000$. By comparison, the smallest amplitude signal of interest in the CMB power spectrum is the kSZ signal, and it is expected to be of the order $\mathcal{D}_\ell \sim 1 \mu {\rm K}^2$ around $\ell = 3000$ \cite{George_2015, Reichardt_2021}. Perhaps, for this reason, the contribution of the CO lines to the CMB power spectrum has been neglected so far in the analysis. 

However, as shown in \cite{Padmanabhan_2018, Keating_2016}, predictions for CO line power spectra from different models differ by more than an order of magnitude. This is mainly due to our limited knowledge of the astrophysics of the galaxies. The main goal of this paper is to revisit this signal and compare the amount of CO line emissions in the CMB maps to other signals/foregrounds at small scales. Using the ``simulated infrared dusty extragalactic sky" (SIDES) simulation \cite{Bethermin_2017, Bethermin_2022}, we show that CO lines have a non-zero correlation with other signals like the CIB and tSZ. Our most important point is that, already at the level of the current CMB data, the total CO signal in combination with these signals is \textit{significant}. In fact, we show that the sum of the cross-power spectra between the CIB and different CO lines can have an amplitude similar to that of the currently estimated kSZ power spectrum level. This result, therefore, paves a new way to measure the CO signal from the CMB data. It may also have a major impact on the potential measurement of the kSZ power spectrum. Our result thus motivates including such signals in all future small-scale CMB data analyses.  

The paper is organized as follows. We first present in Sect.\,\ref{SIDES_smu} the SIDES simulations used to produce, in a consistent way and for the first time, the CO (total and individual lines), CIB, and tSZ maps. The maps are then used to compute the auto- and cross-spectra of these components, which are discussed in Sect.\,\ref{sec:results}. 
This section reveals that the CIB-CO correlation is a bright foreground component i) that has to be considered to revisit the kSZ measurements (Sect.\,\ref{subs:kszmeas}), ii) that can be used to probe the CO signal (Sect.\,\ref{subs:COsignal}) and iii) that has to be included in any cross-correlation analysis (Sect.\,\ref{subs:crosscorr}). The conclusion is given in Sect.\,\ref{sec:cl}. We also discuss in Appendix\,\ref{app:fgxcoind} the contribution of the different rotational CO lines to their cross-correlation with tSZ and CIB.

\section{Maps from SIDES Simulations}\label{SIDES_smu}

SIDES (Simulated Infrared Dusty Extragalactic Sky) is a publicly available\footnote{\url{https://gitlab.lam.fr/mbethermin/sides-public-release}} simulation of the far-infrared (FIR) and sub-millimeter sky based on observed empirical relations \cite{Bethermin_2017, Bethermin_2022}. A detailed description of the simulation is given in \cite{Bethermin_2017, Bethermin_2022, gkogkou2022}. We provide a brief description of the model in the following sections.

\subsection{Continuum emission from dusty star-forming galaxies in SIDES}

The starting point of SIDES is a dark matter lightcone, which is given by the Uchuu N-body simulation \cite{ishiyama2021}. This cosmological simulation achieves both a high mass resolution ($ \rm 3.27 \times 10^8 \, M_{\odot}h^{-1}$) and a large comoving volume (with a comoving side-length box of 2000 $\rm Mpc \: h^{-1}$) between $0 < z < 7$. 
The dark matter halos are populated with galaxies of certain stellar mass through abundance matching. The generated galaxies are then split into passive and star-forming with a probability determined by observations \cite{davidzon2018}. It is assumed that only the star-forming galaxies emit in the FIR and millimeter, so only these types of galaxies need to be assigned with a star formation rate (SFR) value. The SFR values of the main sequence and starburst galaxies are drawn accordingly based on the parameterized fit of the SFR-$M_{\rm stellar}$ relation described in \cite{schreiber2015} taking into account a scatter of 0.3\,dex. The $L_{\rm IR}$ of each galaxy is subsequently defined by the drawn SFR value using the Kennicutt conversion factor ($1.0 \times 10^{-10} M_\odot L_\odot^{-1} {\rm yr}^{-1}$) \cite{kennicutt1998}.  The $L_{\rm IR}$ normalizes the spectral energy distribution (SED) of the galaxy. The shape of the SED depends on the galaxy type (main sequence or starburst) and on the mean intensity of the radiation field ($\langle U \rangle$), which is correlated to the temperature of the dust \citep{magdis2012, bethermin2015, Bethermin_2017}. 

\subsection{CO lines in SIDES}

SIDES also simulates the emission of the strongest high-redshift (sub-)millimeter lines such as [CII], CO, and [CI] \citep{Bethermin_2022}. 
In this work, we focus only on the CO molecular lines. However, other high redshift lines like [CI] are potentially present in the CMB maps as well and should be studied similarly. For CO, the fundamental transition is modeled in SIDES from the $L_{\rm IR}-L_{\rm CO(1-0)}^{'}$ correlation \citep{sargent2014} for the main sequence galaxies, while for the starburst systems there is an offset of -0.46\,dex for $L_{\rm IR}$ at a given $ L_{\rm CO(1-0)}^{'}$. The flux of the other CO transitions is computed using a clumpy and diffuse spectral line energy distribution (SLED) template from \cite{bournaud2015} and following the empirical relation presented in \cite{daddi2015} which connects the flux ratio of CO(5-4) and CO(2-1) transitions with the mean intensity of the radiation field $\langle U \rangle$.\\ 

\subsection{Astrophysical models for tSZ from SIDES halos}
We apply a semi-analytic prescription for the thermal Sunyaev-Zel'dovich (tSZ) effect closely following \cite{websky}, in which we paste gas pressure profiles obtained from detailed hydrodynamic simulations onto the halo catalog obtained from the Uchuu N-body simulation \citep[i.e. as in ][]{sehgal2009, trac2011, osatonagai2022}. The tSZ induces a spectral distortion in CMB photons proportional to the line-of-sight integral of the ratio of electron thermal and rest energies,
\begin{equation}
\Delta T(\nu) = T_{\mathrm{CMB}} \, f(x) \int n_e \frac{k_B T_e}{m_e c^2} \sigma_T dl \equiv T_{\mathrm{CMB}} \, f(x) \, y.
\end{equation}
Here $x = h \nu / k_B T_{\mathrm{CMB}}$ and the frequency dependence relative to the CMB blackbody in the non-relativistic limit is
\begin{equation}
f(x) = \left( x \frac{e^x + 1}{e^x - 1}  - 4 \right).
\end{equation}
The evolution of cluster gas is governed by the physics of star formation and AGN feedback, and as in \cite{websky} we include these effects through a parametric model for the radial cluster pressure profile based on halo mass and redshift, calibrated from detailed hydrodynamic simulations \citep{BBPS2012}. We apply the profile with parameters from Table 1 of \cite{BBPS2012} corresponding to AGN feedback $\Delta = 200$. These are implemented in a publicly available script using the package \texttt{XGPaint.jl}\footnote{\url{https://github.com/xzackli/tsz_sides}}.

\subsection{Why use SIDES?}
The SIDES model is in very good agreement with all the observations we can find for dusty star-forming galaxies (DSFG) in the whole redshift range of SIDES ($0 < z < 7$).  In addition, the model also reproduces more general quantities, such as the stellar mass function (SMF), the evolution of sSFR \cite{Bethermin_2017}. For the purpose of this paper (CO and CIB), the specific observations SIDES reproduces include:
\begin{itemize}
\item The CO line luminosity functions that are observed in the redshift range $1 < z < 3.4$ (Fig. 5 and 6 of \cite{gkogkou2022}).
\item  The measurement of the shot noise level of the CO emission at 3\,mm (Fig. 18 of \cite{Bethermin_2022}).
\item The power spectra of the CIB anisotropies  (Fig. 3 and 4 of \cite{gkogkou2022}).
\item The number counts from Spitzer, Herschel and (sub-)millimeter experiments (e.g. SCUBA2, ALMA), galaxy redshift distributions, number counts per redshift slice and P(D) from SPIRE/Herschel \cite{Bethermin_2017}.
\item  The obscured star formation rate density (SFRD) measurements up to z=4 (Fig. 3 of \cite{Bethermin_2022}). Notice that the observational constraints at $z > 4$ are very dispersed.
\end{itemize}

For tSZ, as discussed in the next section,  the power spectrum from our simulation is compatible (within 1$\sigma$) with the recent measurements.\\

Thus SIDES produces consistent CIB, CO maps, and tSZ maps, both for their emission signals and clustering properties on $\sim$120 square degrees. It is then easy to compute the cross-power spectra to estimate their contribution 
to the small-scale CMB anisotropy measurements. These are the first such simulations consistently containing all these components. 


\section{Results} \label{sec:results}

\subsection{\label{subs:clfgstandard} Comparison between different foreground power spectra}

In this work, we simulate all the components as they would be observed using the SPT telescope. For the CIB emission, we bandpass the galaxy SEDs with SPT-SZ filters at 150 and 220 GHz. On the other hand, to go from frequency independent y map to the tSZ emission corresponding to the 150 and 220 GHz SPT-SZ bandpass, we multiply it by the numbers provided in Tab.~A3 of \cite{Lagache_2020} (which are -0.416 and 9.44 respectively). To generate the CO emission maps, we use an approximate (but fast) method. We assume that the SPT-SZ bandpass at 150 and 220 GHz has uniform spectral response between 125-175 and 195-245 GHz respectively. With this, we then simulate the CO map centered at 150 and 220 GHz with a spectral bandwidth of $\Delta \nu = 50$ GHz. We checked on a 2 Sq. Deg. simulated patch using the measured 150\,GHz SPT-SZ bandpass that this approximation has no impact on the CO power spectra. 
In this section, we present the power spectra of these signal maps and their cross-correlations. It is important to note in our model that the CIB maps include all components of dusty star-forming galaxies, i.e. 1- halo and 2-halo terms (usually called CIB), and shot noise. We also do not implement any flux cut in our CIB maps.

\begin{figure}
\centering
\includegraphics[width=\columnwidth]{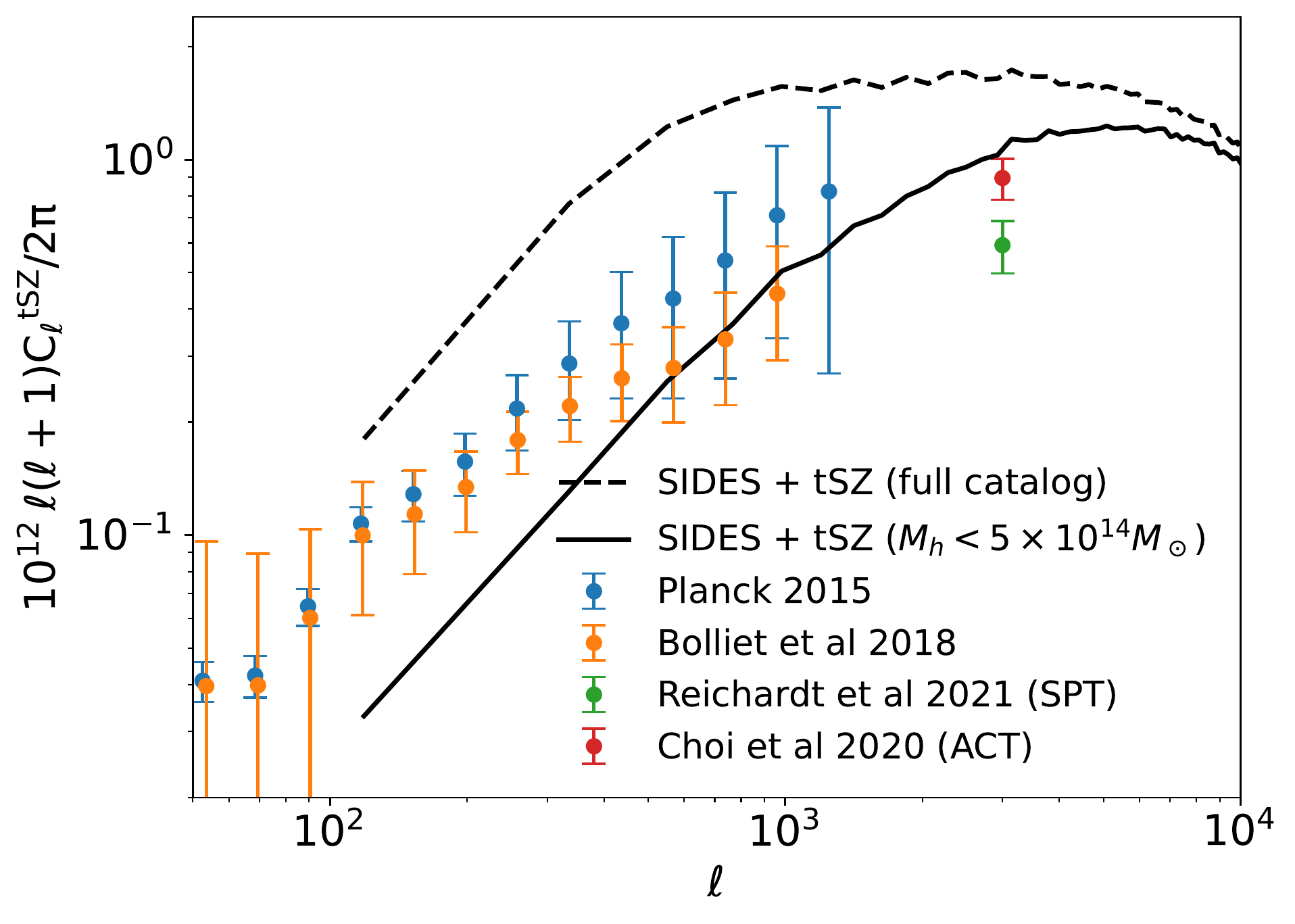}
\centering 
\caption{
tSZ (y) power spectrum from SIDES catalog (black) along with data from Planck \cite{Plancktsz_2016, Bolliet_2018} (blue and orange), SPT \cite{Reichardt_2021} (green), and ACT \cite{Choi_2020} (red) experiments after removing 40 halos massive than $5 \times 10^{14} M_\odot$. 
With this choice, the power spectrum from our simulation is compatible with these measurements for $\ell > 400$, i.e. in the range of scales of interest. 
If these halos are not removed, tSZ power spectrum amplitude from our simulations is much higher than the observations. This is due to the sample variance caused by the small volume of our simulations. }
\label{fig:sides_tsz}
\end{figure}

\begin{figure*}[ht!]
\centering
\includegraphics[width=2\columnwidth]{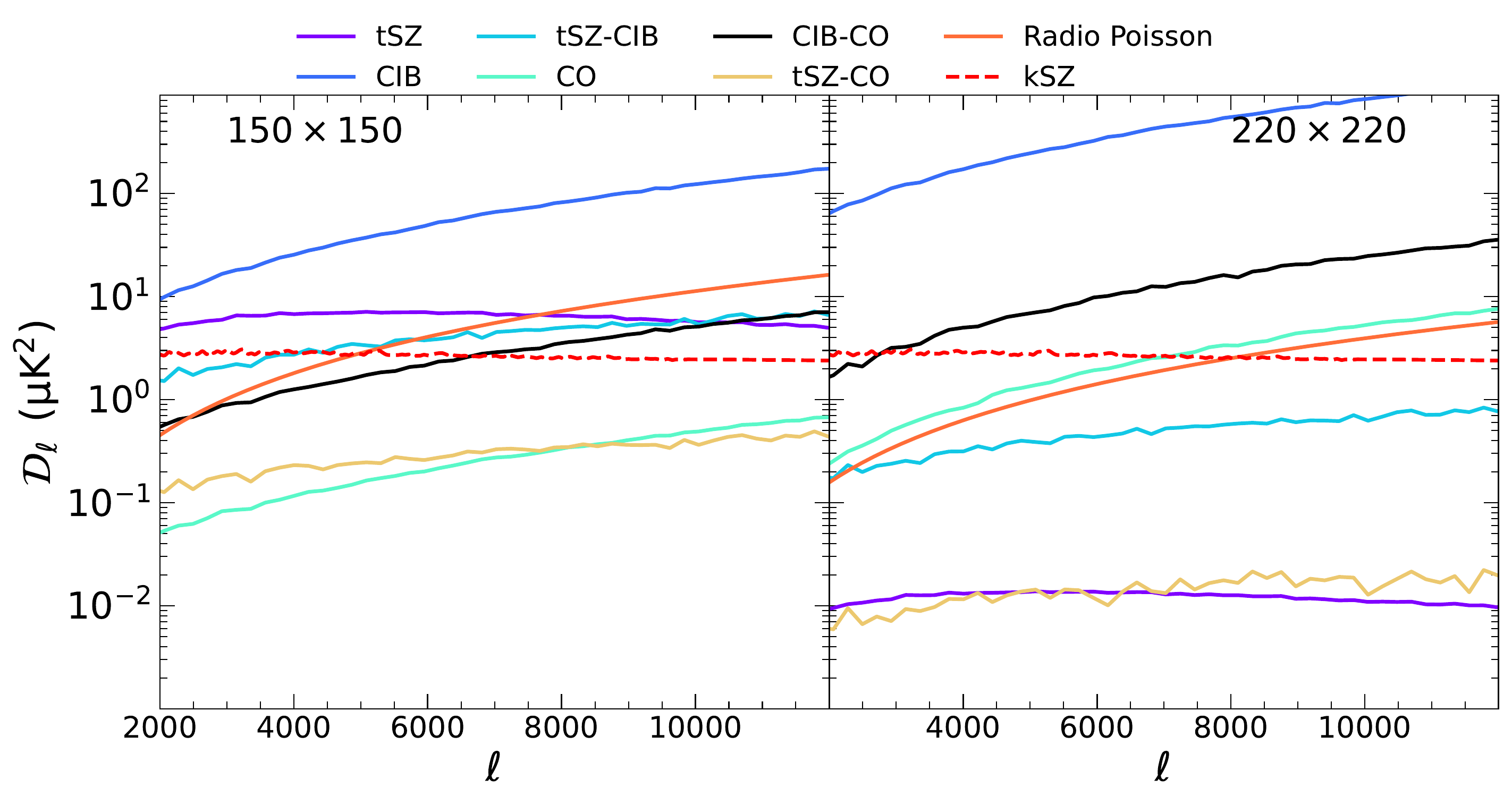}
\centering 
\caption{Power spectrum (absolute value) of different components of the SIDES simulation for SPT 150 and 220 GHz channels. The curves for kSZ and radio components have been taken from \cite{Reichardt_2021}. The CIB-labeled curves include CIB and Poisson. As can be seen, CIB-CO cross-correlation is quite significant at both frequencies. In fact, at 220 GHz, CIB-CO correlation is the second most important signal after the CIB.}
\label{fig:cell_allcomp}
\end{figure*}

First, we verify that our tSZ profile painting method on the dark matter halos works correctly. For this, we plot the frequency-independent power spectrum of our tSZ maps and compare it with the corresponding Planck \cite{Plancktsz_2016, Bolliet_2018}, SPT \cite{Reichardt_2021}, and ACT \cite{Choi_2020} data in Fig.~\ref{fig:sides_tsz} in dashed black curve. We can see that on the relevant small scales here, the tSZ power spectrum from our maps is higher than the measured values. 
Due to its mass weighting, the tSZ power spectrum is very sensitive to Poisson fluctuations \cite{Komatsu_2002}. The high signal seen here is because of the sample variance due to the small volume of our simulation ($\sim 120$ square degrees). After removing halos more massive than $5 \times 10^{14} M_\odot$, we get a tSZ power spectrum more compatible with the observations (solid black line) \footnote{The code used to produce this tSZ map has been extensively tested including reproducing the Websky \cite{Stein_2020} simulations.}. Therefore, we use this catalog with 40 halos $5 \times 10^{14} M_\odot$ removed. We have checked that this choice has insignificant impact on the CIB and CO power spectra as most of the CIB and CO emission comes from halos with masses between $10^{12} - 10^{13} M_\odot$ \cite{Maniyar_2021}. We then proceed to calculate the auto- and cross-power spectra of the CIB, tSZ, and CO lines. This is shown in Fig.~\ref{fig:cell_allcomp} where we plot the power spectrum of all the signals at 150 and 220\,GHz in the left and right panels respectively for SPT. Please note that the `Radio' and `kSZ' power spectra which we do not simulate here are taken from the SPT results from \cite{Reichardt_2021}. 
As expected, the CIB anisotropy power spectrum (shown in blue and including the shot noise) dominates over other signals on such small scales with an amplitude of $\mathcal{D}_{3000}^{\rm CIB} \sim 17 \: {\rm \mu K}^2$ at 150\,GHz. The next dominant component at 150\,GHz is the tSZ power spectrum (violet) with $\mathcal{D}_{3000}^{\rm tSZ} \sim 6.15 \: {\rm \mu K}^2$. Since both the CIB and tSZ are excellent tracers of the large-scale structure (LSS) of the Universe, they are expected to be correlated \cite{Maniyar_2021}. This is exactly what we find here with the CIB-tSZ cross-correlation (cyan) amplitude $\mathcal{D}_{3000}^{\rm tSZ \times CIB} \sim -2.10 \: {\rm \mu K}^2$ at 150\,GHz. We plot the absolute value of the CIB-tSZ cross-correlation since it is negative at 150\,GHz. It is worth noting that the SPT collaboration reported the corresponding values for the total CIB, tSZ, and CIB-tSZ power spectra to be $\mathcal{D}_{3000} \sim 12, 3.42, -0.9 \: {\rm \mu K}^2$ respectively for 150\,GHz. At 220\,GHz, we get a CIB power spectrum amplitude of $115 \: {\rm \mu K}^2$ where SPT collaboration find a value of $121.3 \: {\rm \mu K}^2$. As expected, since the 220\,GHz channel is at the center of tSZ null, both the tSZ and its cross-power spectrum with the CIB is negligible. Our results are therefore comparable with the SPT results. \\

Although we present our results for 150 and 220\,GHz SPT channels here, similar analysis can be carried out at 95 GHz channel as well as for the cross-power spectra between all these frequencies.

\subsection{CO lines and their cross-correlations}\label{subs:clcocross} 
Here we discuss the power spectrum of the CO lines and their cross-correlations with other signals. Instead of showing the individual CO($J \rightarrow J-1$) lines, in Fig.~\ref{fig:cell_allcomp}, we plot the sum of their power spectra for clarity. As can be seen, at $\ell = 3000$ the sum of all the CO line power spectra, shown in blue, is $\mathcal{D}_{3000}^{\rm CO, total} \sim 0.08 \: {\rm \mu K}^2$ at 150 GHz. In comparison to the other signals, CO power spectrum is at least an order of magnitude smaller at 150\,GHz. For 220\,GHz, we find that the total CO power spectrum has the same amplitude as that of the radio component. However, at $\ell = 3000$, they are both much smaller than the CIB and kSZ power. This therefore supports the previous CMB power spectrum analysis studies which did not account for the presence of the CO lines in the CMB maps. \\

However, things get interesting when we consider the cross-correlations between CO and the CIB and tSZ. CO lines are tracers of star-forming galaxies (and hence the LSS) similar to the CIB. They are therefore expected to be correlated with the CIB and tSZ. 
This is exactly what we find. For the CIB-CO cross-power spectrum, we get $\mathcal{D}_{3000}^{\rm CIB \times CO, total} \sim 0.89 \: {\rm \mu K}^2$ at 150\,GHz. The CIB-CO cross-correlation has therefore similar power as (absolute) CIB-tSZ cross-correlation at 150\,GHz. On the other hand, we get $\mathcal{D}_{3000}^{\rm CIB \times CO, total} \sim 3.20 \: {\rm \mu K}^2$ at 220\,GHz making it the biggest signal after the CIB. This shows that although we can neglect the auto-power spectrum of different CO lines due to their very small amplitude, their cross-correlation with the CIB can be large and significant already for the current data.

\begin{figure}
\centering
\includegraphics[width=\columnwidth]{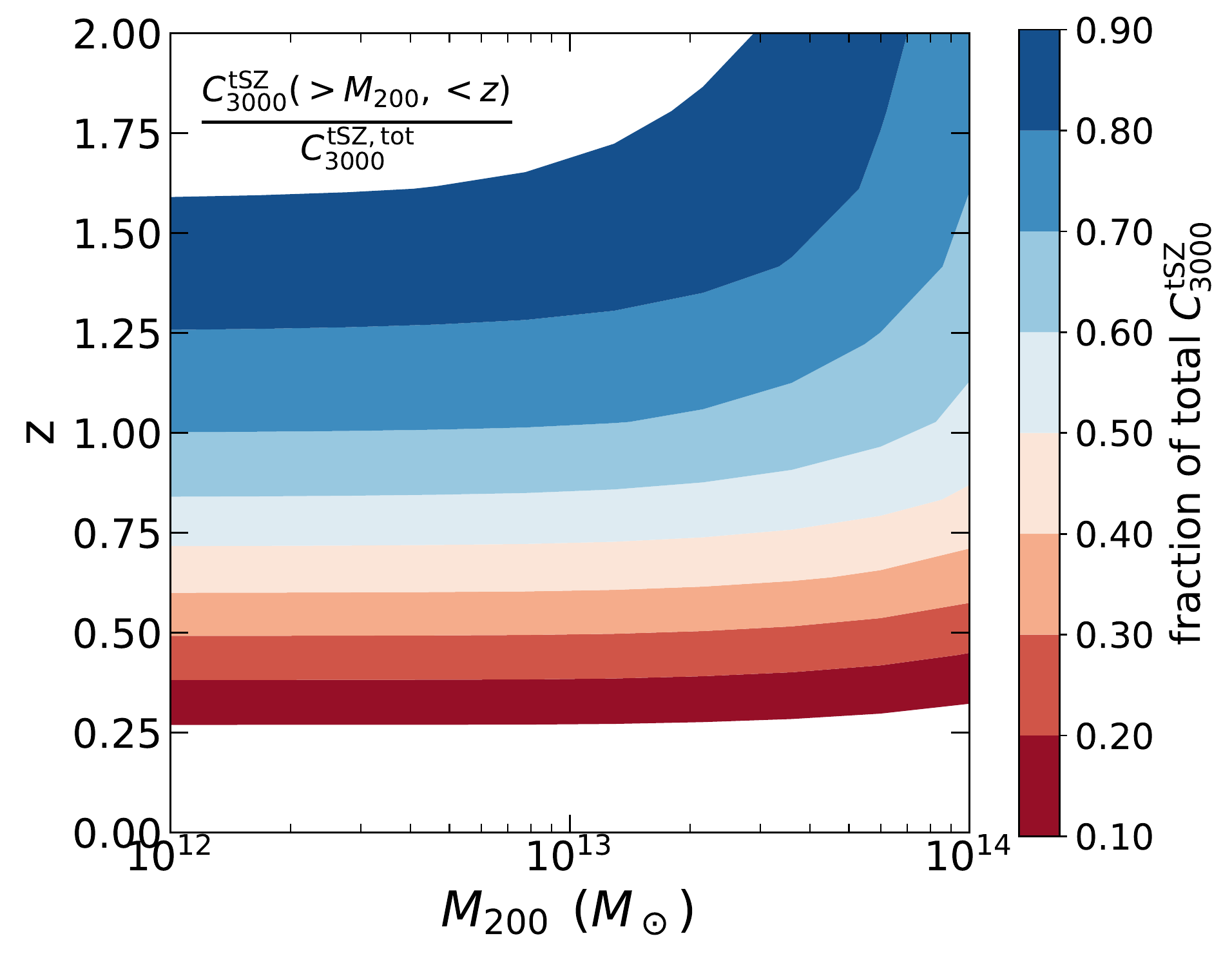}
\centering 
\caption{Two-dimension contour plot showing the fraction ($10\%$ to $90\%$) of the total tSZ power at $\ell = 3000$ as a function of different mass and redshift ranges. This shows that the majority ($\sim 70\%$) of the tSZ power is sourced by massive halos $M_{200} > 10^{13}M_\odot$ at low redshifts $z < 1$.} 
\label{fig:tsz2d}
\end{figure}

On the other hand, although the tSZ signal is quite large on these scales, we find that the cross-correlation between the tSZ and CO lines is small and not as important as the CIB-CO correlation. This is expected at 220\,GHz where the tSZ signal is itself negligible. At 150\,GHz, its amplitude $\mathcal{D}_{3000}^{\rm tSZ \times CO, total} \sim -0.18  \: {\rm \mu K}^2$ is similar to the total CO power spectrum. In order to understand this, we explore the origin of the tSZ power. Similar to \cite{Bhattacharya_2012}, we plot the fractional tSZ power for halos with mass $> M_{200}$ and redshift $< z$, from 10-90\% of the total power spectrum from our simulation in Fig.~\ref{fig:tsz2d}. This plot shows that majority of the tSZ power ($\sim 70\%$) comes from massive halos ($M_{200} > 10^{13} M_\odot$) at low redshifts ($z < 1$). 

\begin{figure}
\centering
\includegraphics[width=\columnwidth]{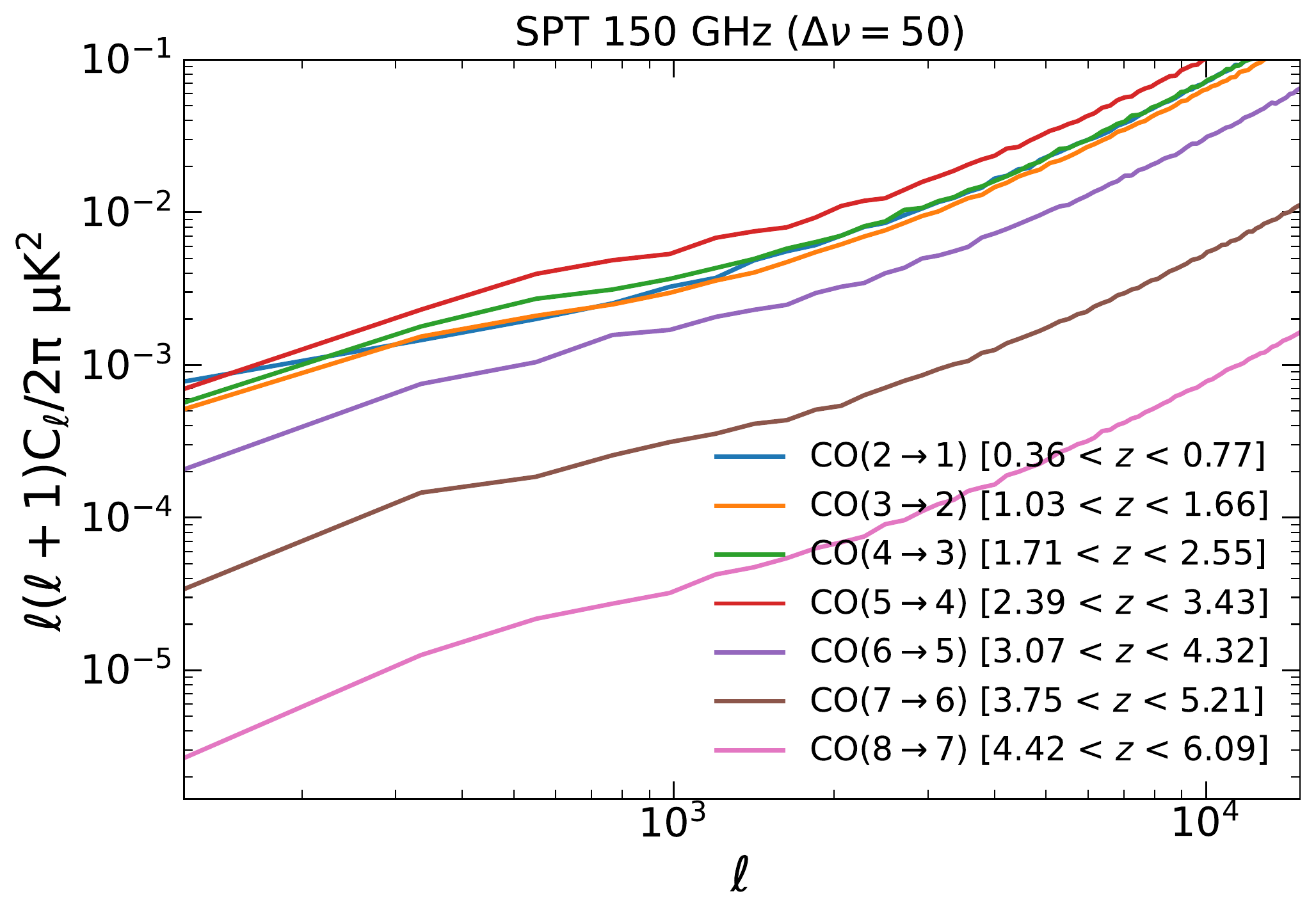}
\centering 
\caption{Power spectra of individual CO lines observed in the SPT 150\,GHz maps. Most of the power comes from the different CO lines between $1 < z < 3.5$.}
\label{fig:COind}
\end{figure}

As we precisely know the rest frame frequency of different CO($J \rightarrow J-1$) transitions, we can determine their redshift distribution within the SPT bandpass. Since we have created our CO maps assuming a spectral bandwidth of $\Delta \nu = 50$ centered at 150\,GHz, we have different CO lines at various redshifts which fall within the observed frequencies of $130 < \nu < 170$\,GHz. This means that CO($1 \rightarrow 0$) with the rest frame frequency of 115.27\,GHz, does not show up in our maps. In fact, apart from CO($2 \rightarrow 1$) transitions from $0.35 < z < 0.75$, the rest of the CO lines in our map are from higher redshifts ($z > 1$). In Fig.~\ref{fig:COind}, we show the power spectra of individual CO lines from our maps. As we can see, at $\ell = 3000$, the CO($5 \rightarrow 4$) line ($2.4 < z < 3.4$) has the highest power followed by CO($4 \rightarrow 3$) line ($1.7 < z < 2.5$), CO($2 \rightarrow 1$) line ($0.35 < z < 0.75$), and CO($3 \rightarrow 2$) line ($1.0 < z < 1.7$) which have almost the same power. Therefore, most of the CO power comes from redshifts $z > 1$. All these factors help us understand the smaller amplitude of the tSZ-CO cross-correlation compared to the CIB-CO cross-correlation for the SPT 150\,GHz channel. 

In Appendix~\ref{app:fgxcoind}, we show the cross-correlation of individual CO lines with the CIB and tSZ in Fig.~\ref{fig:cibxCOind} and Fig.~\ref{fig:tszxCOind} respectively and discuss their implications.

\section{Implications of this new foreground}

\subsection{Implications for kSZ measurements}\label{subs:kszmeas}
Measuring the kSZ power spectrum from the data is an important goal for the current and upcoming CMB experiments. The SPT collaboration has performed multi-component analyses of the small scale CMB power spectrum accounting for different foregrounds \cite{George_2015, Reichardt_2021}. In fact, their recent analysis reported a first measurement of the kSZ power at $\geq 3\sigma$ with a value of $\mathcal{D}_{3000}^{\rm kSZ} = 3.0 \pm 1.0 \: {\rm \mu K}^2$ (Fig.~\ref{fig:cell_allcomp}). They also show that the kSZ power is degenerate with the corresponding tSZ and CIB-tSZ cross-correlation power respectively. In this analysis, they do not account for the presence of CO lines in the SPT maps. As shown in Sect.~\ref{subs:clcocross}, the total auto-power spectrum of the CO lines is an order of magnitude smaller ($\sim 0.1 \: {\rm \mu K}^2$) than the expected kSZ power spectrum ($\sim 1 \: {\rm \mu K}^2$), so this should not affect the results. However, here we have shown that the cross-correlation of the CIB with CO lines has an amplitude similar to the kSZ and CIB-tSZ cross-correlation. Looking only at the 150\,GHz, a significant (negative) cross-correlation between the tSZ and CO lines could have resulted in a fortuitous cancellation of the CIB-CO correlation, which is not the case here. This is a very important result and motivates current and future analyses to include this CIB-CO correlation component. 

Since this new signal has almost the same power as the kSZ, it may significantly bring down the measured kSZ power. On the other hand, the shape of this CIB-CO correlation seems similar to the other signals such as the CIB and CIB-tSZ correlation. This is different than the kSZ which is mostly flat on these scales. It is thus possible that without affecting the kSZ power, this new signal may change values of the CIB and CIB-tSZ correlation (but also other components as radio). Without performing a thorough multi-component analysis, it is therefore not possible to say whether this will decrease the amount or signal-to-noise ratio of the kSZ power detected by the SPT team.
In an upcoming paper, we will perform such an analysis to determine the impact of CIB-CO correlations on kSZ power spectrum.

\subsection{CO as a signal}\label{subs:COsignal}

CO($J \rightarrow J-1$) molecular line emissions are among the brightest lines in the galaxy spectra. Their measurements can be used to estimate the amount of gas mass and star formation rate in galaxies \cite{Bernal_2022}. They also act as a proxy for cosmic molecular gas history. Thus CO lines are one of the prime targets of some of the current and upcoming LIM surveys. Recently, \cite{Keating_2016} and \cite{Keating_2020} reported a $2\sigma$ and $4\sigma$ detection of the shot-noise component of the CO power spectrum. However, to date, there has been no detection of the clustering component of the CO power spectrum. As we showed in the previous section, the CIB-CO correlation acts as a significant signal in the small-scale CMB data. Thus, those data can be potentially used to measure the CIB-CO cross-correlation signal for the first time.

As mentioned in Sect.~\ref{sec:intro}, predictions for CO power spectra can vary by a large amount between different models. Such a measurement will therefore not only help in ruling out some existing models but also constrain the parameter space of the most promising models. This will, in turn, enable us to draw conclusions on e.g. the cosmic molecular gas history. As several CO lines from different redshifts show up at different frequencies in the CMB maps, measuring this amplitude across these frequency channels has the potential to break some potential degeneracies in the models.

\subsection{Implications for studies involving cross-correlations}\label{subs:crosscorr}
In view of the high quality data from the current and upcoming cosmological experiments, cross-correlations of different tracers of the LSS have emerged as a powerful tool to get rid of different biases and isolate signals of interest. All such analyses involving small-scale CMB information will be affected by this new signal. One example is the so-called projected-field estimator to detect the kSZ signal \cite{Ferraro_2018}. It involves squaring the CMB map and cross-correlating it with a LSS tracer or in general measuring the $\langle TT\delta \rangle$ three-point function where $T$ is the CMB data and $\delta$ is a LSS tracer. Such analyses will be affected by the CO lines present in the CMB maps as they will cross-correlate with the LSS tracer and CIB, and therefore act as a source of bias (as $\langle {\rm CO}-{\rm CO}-\delta \rangle$ and $\langle {\rm CO}-{\rm CIB}-\delta \rangle$) terms to the kSZ signal if not properly accounted for. 
In fact, this could be one of the components responsible for an unexplained excess kSZ amplitude observed by \cite{Kusiack_2021} using this estimator. Therefore, with the more precise CMB data, CO emission in galaxies may give non-significant contributions and thus must be accounted for in cross-correlation analyses. 

\section{Conclusion} \label{sec:cl}
As high-resolution CMB data keeps on getting more precise with the current and upcoming experiments, this presents us with an exciting opportunity to perform a wide range of analyses. In order to best utilize this rich data, it is therefore crucial to understand and account for all the sources contributing at these millimeter frequencies. In this paper, we study the extragalactic CO molecular emission lines in such a context, which traditionally have been neglected in past analyses. 

Since the CMB experiments use broadband filters, we thus expect different CO lines from several redshifts to fall within these bands. Using the SIDES simulations, we prepare a map containing the CIB, tSZ, and CO lines as observed by the SPT telescope at 150 and 220\,GHz. These are the first such simulations consistently containing all these components. We show that our predictions of the CIB, tSZ power spectra, and their cross-correlations match the observations. 
We show that the power spectrum of the CO lines is an order of magnitude smaller than the other foregrounds. Their contribution to the total power spectrum at these scales is thus negligible. However, we then go on to show that the cross-correlation of the CIB with CO lines has a contribution similar to the CIB-tSZ correlation and the kSZ power spectrum. It therefore contributes a non-negligible amount to the total power at these scales. This is the main result of this paper. We also show that, unlike the CIB-CO, the tSZ-CO correlation is not as strong at 150\,GHz. This is due to the difference in the redshift and halo masses of the sources contributing to these two signals. Similar arguments are expected to hold at 95 GHz as well which is the other main CMB channel.\\

Our result is of special significance in light of the recently reported first-ever $>3 \sigma$ detection of the kSZ power spectrum by \cite{Reichardt_2021}. This result is quite important as kSZ measurements and their interrogations promise to shed light on the reionization of the Universe. They perform this measurement of the kSZ accounting for the presence of several other foregrounds like the CIB, tSZ, and radio emission. However, they do not account for the CO lines and their cross-correlations with the CIB which we show here have an amplitude similar to the kSZ power spectrum. As discussed in Sect.~\ref{subs:kszmeas}, our results call for a more careful analysis of the data accounting for CIB-CO cross-correlations, which may have a significant impact on the kSZ measurements. \\

Our results also show that the CO lines present in the CMB maps will be crucial for all the small scale CMB auto-power spectrum and cross-correlation studies involving a LSS tracer. Upcoming CMB experiments aim to use small scale CMB measurements to constrain a diverse range of physics including precision constraints on the $\Lambda$CDM model to neutrino mass to decaying dark matter \citep[e.g.][]{Abazajian_16}. The signatures of many of these processes are very subtle and a preeminent challenge is mitigating possible biases. Not modelling the CO lines risks their power being incorrectly attributed to novel physics. Hence, including the CO lines in these analysis is essential to avoid any potential biases. \\

In an upcoming paper, we will perform a comprehensive study of the small-scale CMB power spectrum with all these foregrounds. This analysis will not only determine the impact of CIB-CO correlations on kSZ power spectrum measurements but will also provide us with the first CIB-CO cross-correlation measurement from the CMB maps. This measurement will help us constrain the star formation history of the Universe as both the CIB and CO trace star-forming galaxies within our Universe. 

Finally, in this paper, we solely focused on the extragalactic CO molecular lines. There are other lines like [CI] contributing to the millimeter sky \cite{Bethermin_2022}. As the data quality keeps on getting better, we might have to look beyond CO lines and also account for other lines like [CI].

\acknowledgments

We thank Srinivasan Raghunathan from the SPT collaboration for providing us with the SPT measurements of the Radio and kSZ power spectrum. We thank Matthieu B\'ethermin for his help in using SIDES. This project has received funding from the European Research Council (ERC) under the European Union’s Horizon 2020 research and innovation program (grant agreement No 788212) and from the Excellence Initiative of Aix-Marseille University-A*Midex, a French ``Investissements d’Avenir" program. Some of the results in this paper have been derived using various standard Python packages (numpy, scipy, matplotlib).

Part of this work was carried out on the ancestral land of the Muwekma Ohlone Tribe, successors of the Verona Band of Alameda County.

\bibliographystyle{prsty.bst}
\bibliography{refs}

\newpage
\onecolumngrid
\appendix

\section{Correlations of individual CO lines with foregrounds}
\label{app:fgxcoind}

\begin{figure}[!h]
\centering
\includegraphics[width=0.7\columnwidth]{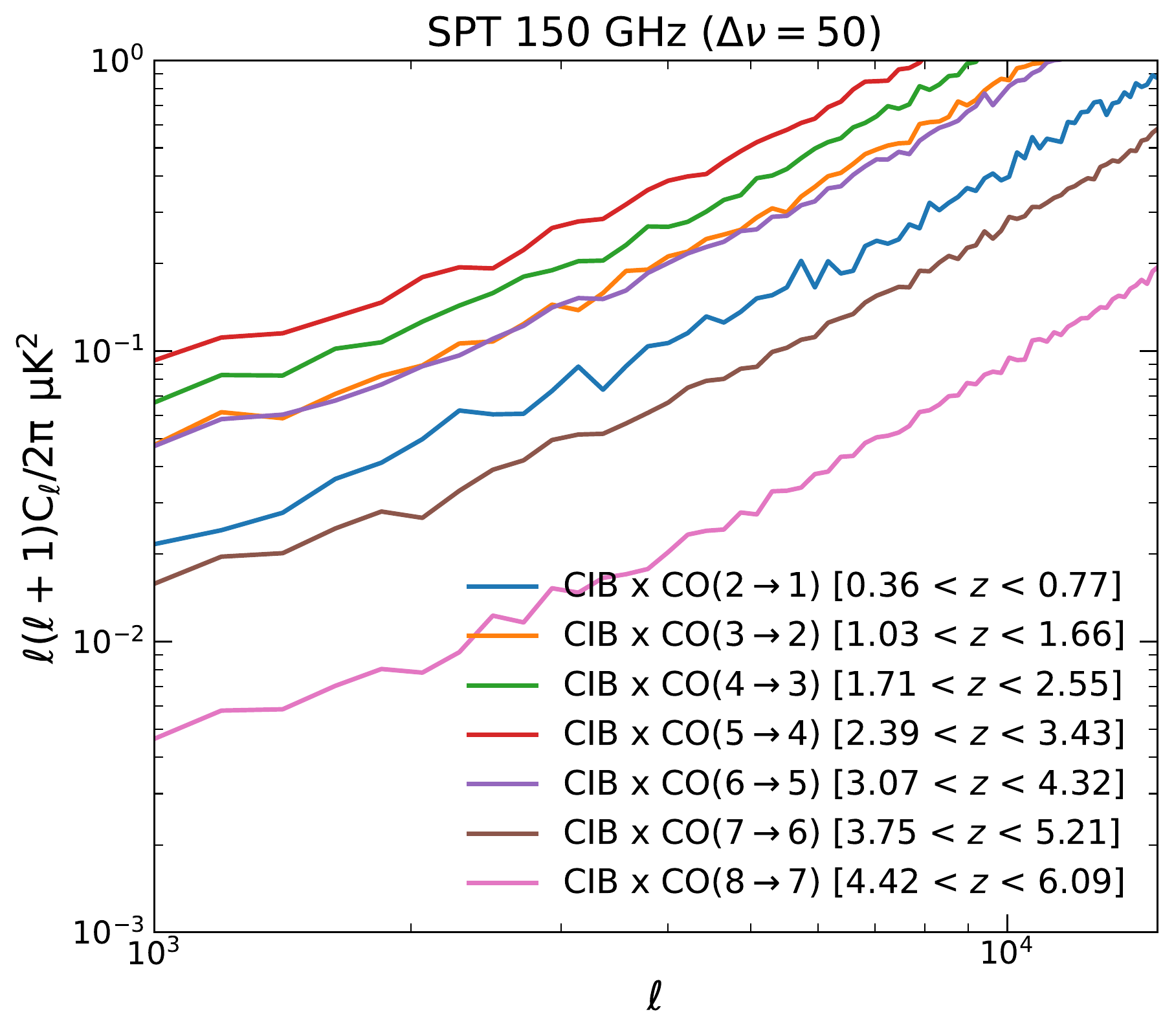}
\centering 
\caption{Cross-power spectra of individual CO lines with the CIB at 150\,GHz.}
\label{fig:cibxCOind}
\end{figure}

Fig.~\ref{fig:cibxCOind} shows the cross-power spectra between different CO lines and the CIB at 150 GHz. The correlation is highest for CO($5\rightarrow4$) line from $2.4 < z < 3.4$. This is followed by CO($4\rightarrow3$) and, CO($3\rightarrow2$) lines. This is similar to the individual CO line power spectra plotted in Fig.~\ref{fig:COind}. However, in Fig.~\ref{fig:COind}, CO($2\rightarrow1$) has the same amplitude as that of CO($3\rightarrow2$) which is not the case here where CIB-CO($3\rightarrow2$) correlation is much higher than CIB-CO($2\rightarrow1$). This can be explained as follows.
As the CIB is traced by star-forming galaxies, bulk of the CIB emission comes from the epoch when the star formation in the Universe peaked between $1 < z < 3$ \cite{Maniyar_2018}. Also, the CIB has this peculiar property that low frequency CIB maps trace galaxies at higher redshifts and vice-versa. This also explains higher correlation of the CIB with the CO($3\rightarrow2$) line than CO($2\rightarrow1$).

\begin{figure}
\centering
\includegraphics[width=0.7\columnwidth]{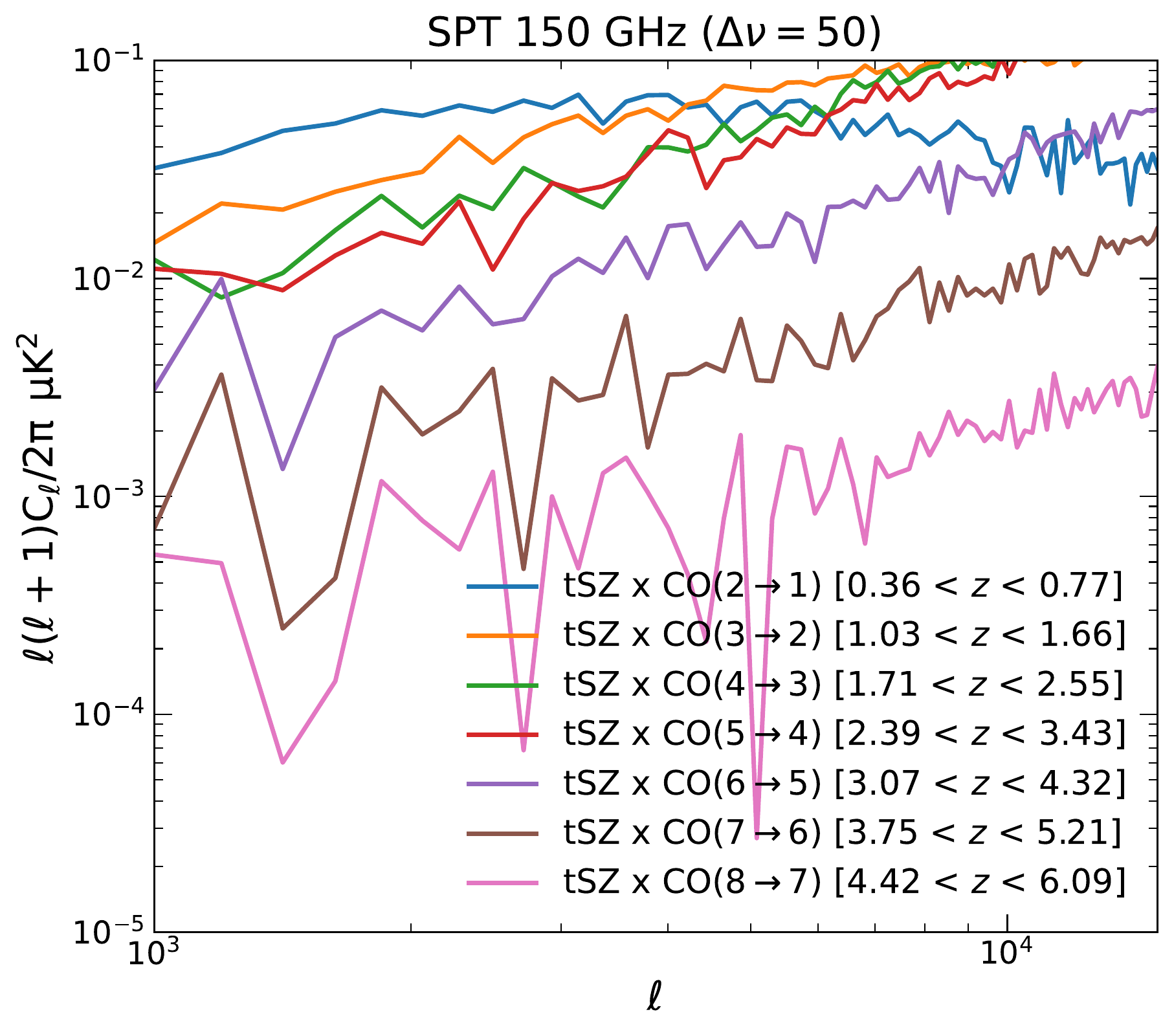}
\centering 
\caption{Cross-power spectra of individual CO lines with the tSZ at 150\,GHz.}
\label{fig:tszxCOind}
\end{figure}

On the other hand, as can be seen from Fig.~\ref{fig:tszxCOind}, this is not the case for tSZ-CO correlations. Although CO($5\rightarrow4$) line has the highest auto-power spectrum at 150 GHz, tSZ has the highest correlation with CO($2\rightarrow1$) line originating between $0.4 < z < 0.8$. As explained in Sect.~\ref{subs:clcocross}, this is due to the low redshift origin of most of the tSZ power. In fact, we can see from Fig.~\ref{fig:tsz2d} that most of the tSZ power comes from massive halos at low redshifts ($z < 1$). Therefore, as the higher CO transition lines trace higher redshifts, their correlation with the tSZ keeps going down as well.

\end{document}